\documentclass[11pt,twoside]{article}
\usepackage{./asp2010}

\resetcounters

\begin{document}

\title{Spectroscopy and Interferometry of Luminous Blue Variables}
\author{N. D. Richardson$^1$, D. R. Gies$^1$, N. D. Morrison$^2$, G. Schaefer$^3$, T. ten Brummelaar$^3$, J. D. Monnier$^4$, \& J. R. Parks$^1$
\affil{$^1$Center for High Angular Resolution Astronomy, Department of Physics and Astronomy, Georgia State University, P. O. Box 4106, Atlanta, GA  30302-4106; richardson@chara.gsu.edu}
\affil{$^2$Ritter Astrophysical Research Center, Department of Physics and Astronomy, University of Toledo, 2801 W. Bancroft, Toledo, OH 43606}
\affil{$^3$The CHARA Array of Georgia State University, Mt.~Wilson, CA}
\affil{$^4$Department of Astronomy, University of Michigan, Ann Arbor, MI 48104}}

\begin{abstract}
We report on the preliminary results of a three-year monitoring campaign of Galactic and Magellanic LBVs with both moderate and high resolution spectroscopy. We have collected more than 500 moderate-resolution spectra of 20 stars in the red portion of the optical spectrum, which includes the wind-sensitive transitions of H$\alpha$ and He I 5876 and 6678 Angstroms. We summarize our detailed study of 15 years of high resolution H$\alpha$ spectroscopy of the prototypical luminous blue variable, P Cygni. We report on the discovery of discrete absorption components in P Cygni's H$\alpha$ profile, and we discuss their relationship to structure in the wind. These results are compared to our recent high resolution interferometric imaging of the $H$-band emitting region surrounding the star. We discuss recent observations of $\eta$ Car, highlighting its unusual recovery from the previous periastron passage. Our results on HDE 326823 indicate that the star has a short period binary orbit, with Roche Lobe overflow onto an unseen massive companion.

\end{abstract}

\section{Introduction}

Luminous Blue Variables (LBVs) are interesting massive stars that have presumably exhausted their supply of hydrogen in their cores. They have unusually high mass loss rates ranging from $\approx 10^{-6} M_{\odot}$ yr$^{-1}$ to $10^{-3} M_{\odot}$ yr$^{-1}$, and can experience great eruptions. The typical long-term variability of these stars involves a change of their observed spectral types from O, at minimal optical flux, to that of an A or F supergiant at maximal optical flux, keeping the bolometric luminosity roughly constant. These excursions will typically last a few years and the mass loss rates will change throughout the process. Unfortunately, little systematic monitoring of the known population has been performed, leaving the driving mechanism(s) for this long term variability a mystery. 

We began a systematic monitoring campaign of roughly 20 LBVs using the CTIO 1.5 m telescope and its Cassegrain spectrograph ($R \sim 3,000$) to probe the physics of the long-term variability for Galactic and Magellanic LBVs. The main criterion for an LBV to be considered part of this survey was the observed optical flux because we needed to obtain reasonable S/N in the continuum in a short amount of time. As a result, we have obtained data on all Galactic LBVs and candidate LBVs brighter than $V=11$ and located south of $\delta = + 20^{\circ}$. Selected targets, such as $\eta$ Carinae, were observed with the echelle spectrograph ($R \simeq 40,000)$ on the same telescope. In addition, we have obtained high resolution spectroscopy on P Cygni using Ritter Observatory's 1 m telescope and echelle spectrograph ($R \simeq 26,000$). We have some interferometric observations to confirm the results from spectroscopy through direct imaging of these stars on the milliarcsecond scale. We present the average spectrum of each of our Galactic targets in Figure 1, where we sort the targets by peak H$\alpha$ height. In general, mass loss rates increase in the vertical direction, and range from $\sim 10^{-6} M_{\odot}$ yr$^{-1}$ to $10^{-3} M_{\odot}$ yr$^{-1}$. Quantitative efforts are underway to characterize how variations in the mass loss rate change the observed spectrum. Here we will describe some results on specific targets, including P Cygni, $\eta$ Carinae, and HDE 326823. We will also present some preliminary results from the full survey.

\begin{figure} 
\begin{center} 
\includegraphics[angle=0, height=12cm]{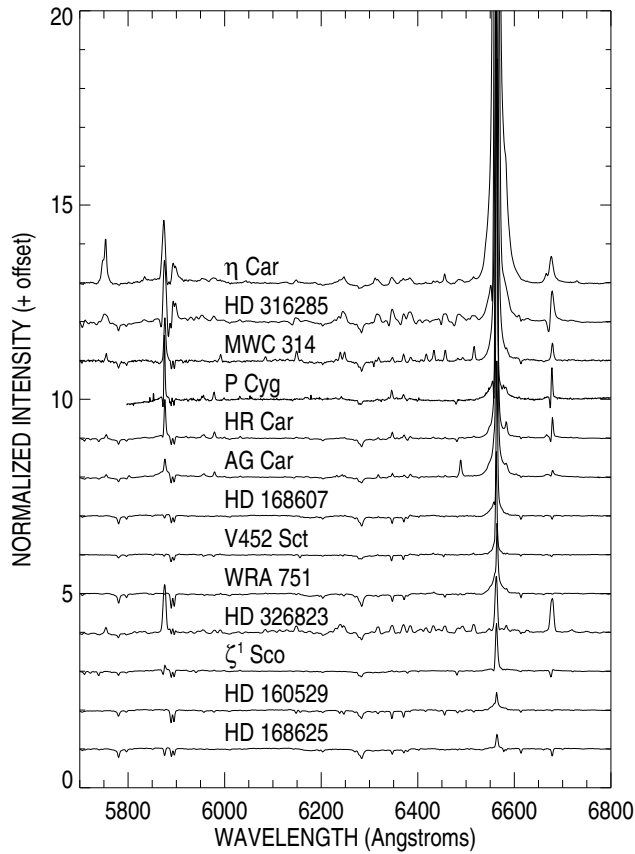}
\end{center} 
\caption{Spectra of Galactic LBVs and LBV candidates from our monitoring program. Spectra are sorted by peak H$\alpha$ emission and offset for clarity.} 
\label{fig1} 
\end{figure} 
\clearpage


\section{P Cygni}

We recently presented (Richardson et al.~2011) results on a long-term program monitoring the photometric and H$\alpha$ variability of the prototypical LBV, P Cygni. These results demonstrated the cohesive variability of both the continuum ($V-$band) and H$\alpha$ observables, including net equivalent widths, absorption edge velocities, the kinematic width of the emission, and radial velocity of the profile. In addition, we found that the absorption component of the P Cygni profile had moving sub-features reminiscent of the Discrete Absorption Components (DACs) observed in resonance lines of hot stars. However, unlike typical O star DACs, these had a timescale on the order of one thousand days, and repeated on timescales of 1700 days. In addition, the strength of these DACs were correlated with the net H$\alpha$ profile strength and the $V-$band photometry. We concluded that the star has a nearly spherical wind.

We observed this star and its wind using Georgia State University's CHARA Array (ten Brummelaar et al.~2005) and the Michigan InfraRed Combiner (MIRC; Monnier et al.~2006). This instrument is capable of measuring the observed visibility and closure phase information across eight spectral channels in the $H-$band. These observables yield information about the size and asymmetry of the source. Our initial image reconstruction analysis, shown in Figure 2, shows a nearly spherical shell, with a small ($<1\%$) asymmetry to the north, very close to the star. These results are in good agreement with the spectroscopic results (Richardson et al.~2011). This year, we collected more data in order to provide better constraints on any asymmetry observed near the star and in its outflow. 

\begin{figure} 
\begin{center} 
\includegraphics[angle=0, height=7cm]{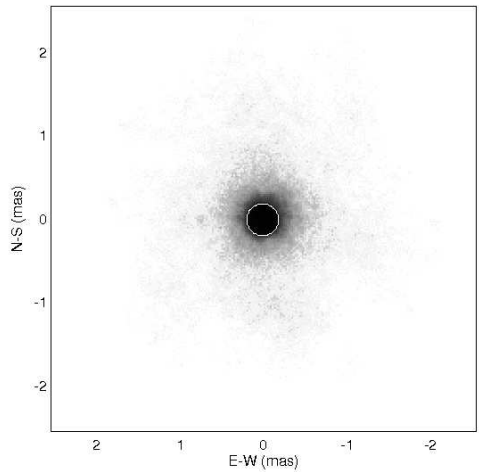}
\end{center} 
\caption{Interferometric image reconstruction of P Cygni from observations collected at CHARA with the MIRC beam combiner. The best fit model includes a star of 0.4 mas diameter (circle), consistent with results of Najarro et al.~(1997), and a Gaussian distribution of wind flux with a FWHM of 1.15 mas.} 
\label{fig2} 
\end{figure} 
\clearpage 

\section{$\eta$ Carinae}

The binary system $\eta$ Carinae experienced major eruptions in the mid nineteenth century that expelled several solar masses of material. It has a long 5.5 yr orbital period and high eccentricity ($e \sim 0.9$). We observed this system almost every night during the 2009.0 periastron passage and demonstrated how the H$\alpha$ profile changes around periastron (Richardson et al.~2010). These results are consistent with those from other campaigns that occurred at that time (see contributions from Damineli, Madura, Teodoro, and Corcoran in these proceedings). In general, there is a large collapse of flux from the system in our line of sight. The narrow lines from the ejecta and the wind lines respond to these changes based on their formation radii. Our follow-up study of other wind lines during the minimum is in preparation and should be completed in the near future.

Mehner et al.~(2010) demonstrated that the wind lines have weakened over the last two binary orbits. We have compared our ground based results to the results from {\it HST/STIS} observations from previous cycles by summing across the entire STIS slit to include nearby nebulosity. Our results also show that the wind lines are weaker than previous observations. Both our new echelle data and those from {\it HST/STIS} demonstrate that the system's mass loss rate is changing. In-depth studies of the system will allow us to better understand the complicated physics in the rare LBV binary systems.

\section{HDE 326823}

HDE 326823 (Hen 3-1330; V1104 Sco; ALS 3918; ASAS J170654-4236.6) is an unusual transition-phase object in the Galaxy. Van Genderen (2001) places this object in the category of ex-/dormant LBVs, and the star is is thought to be entering the WN stage of evolution from either the LBV or RSG stage (e.g., Marcolino et al.~2007). Analysis of the recent photometric variations from the All Sky Automated Survey by Pojma\'{n}ski \& Maciejewski (2004) indicates a short period of 6.123 d and an amplitude of 0.17 mag in $V$. 

We have further analyzed our echelle data on HDE 326823 since the conference, and found that the star's variability is best explained in a binary scenario. We measured radial velocities using cross-correlation for the N II absorption lines in the region of 5667--5710 \AA. The resulting binary orbit has a period of 6.1228 days, an eccentricity of $0.19 \pm 0.06$, and a large mass function, $f(M) = 7.6 M_\odot$. We suggest that the absorption lines originate in a mass donor, while the more massive gainer is enshrouded in an optically thick torus. The mass donor loses mass via Roche Lobe overflow onto the mass gainer, and through the L2 point into a circumbinary disk, the source of the optical emission lines. This model can explain the unusual abundances observed ($X_H < 3\%$; Marcolino et al.~2007) and the lack of kinematic variability of the emission lines. 

\section{Future Directions for the Survey}

We have collected more than 500 moderate resolution red spectra around H$\alpha$ using the CTIO 1.5 m. In addition, we have obtained many low resolution NIR $H$ and $K$ band spectra of LBVs north of $\delta \approx -35^{\circ}$. We are beginning our analysis of these data by using stars with mass loss rates that are known from modeling or radio observations to calibrate a scale for how $\dot M$ affects the red and NIR spectra for these targets. Our results are most consistent if HDE 326823 is excluded, as that star is likely a Roche Lobe Overflow binary, and not a bona fide LBV. These results will allow us to produce estimates for $\dot M$ for many LBVs that have no known estimate for their mass loss rates.

In addition, we plan on searching for new binaries from the sample of stars with a large number of spectra. Many of the cooler (and those with lower $\dot M$) LBVs show photospheric Si II 6347, 6371 absorption lines. With cross correlation, we may be able to find velocity variations associated with orbital motion, with periods as long as three years. We have three candidate stars now, and we plan to obtain high resolution spectroscopy to confirm if these binaries are real. In addition, we have obtained interferometric data with the {\it HST} Fine Guidance Sensors to search for close ($\sim 0.''1$) companions to the LBVs (see the similar study for O stars by Caballero-Nieves in these proceedings). We detect one faint companion, but large distances and potentially large $\triangle V$ may result in fewer detections in the astrometric search for companions. 

Lastly, we will examine the variability of the population, through line profile changes, spectral changes, and photometric changes from data obtained through the ASAS and the AAVSO. Our preliminary results on the LMC prototype, S Dor, show that the star is now the hottest that has been observed in the last century. {\it HST/STIS} observations in the ultraviolet are planned to probe the deep underlying photosphere of this star in order to better understand the photospheres of these objects. The variability of these stars may reveal what the driving mechanism is for the long-term S Doradus type variations. This variability has not been understood theoretically, and observationally, it is the best way to classify an LBV candidate as a bona fide LBV.

\acknowledgements We gratefully acknowledge the support of the observers at Ritter Observatory, staff of the CHARA Array, and the SMARTS observers at the CTIO 1.5 m telescope for their assistance with observations. Some spectra included in Figure 1 were obtained at GSU's Hard Labor Creek Observatory, and we acknowledge Emily Aldoretta, Ben Jenkins, and Nic Scott for their assistance. This material is based upon work supported by the National Science Foundation under Grant No.\ AST-1009080.

\section*{References}
\noindent

Marcolino, W. L. F., de Araujo, F.X., Lorenz-Martins, S., \& Borges Fernandes, M. 2007, AJ, 133, 489

Mehner, A. et al. 2010, ApJ, 717, L22

Monnier, J. D., et al. 2006, ApJ, 647, 444

Najarro, F., Hillier, D. J., \& Stahl, O. 1997, A\&A, 326, 1117

Pojma\'{n}ski, G., \& Maciejewski, G. 2004, Acta Astron., 54, 153

Richardson, N. D., et al. 2010, AJ, 139, 1534

Richardson, N. D., et al. 2011, AJ, 141, 120

ten Brummelaar, T., et al. 2005, ApJ, 628, 453

van Genderen, A. M. 2001, A\&A, 366, 508

\end{document}